\documentclass[twocolumn,showpacs,preprintnumbers,amsmath,amssymb]{revtex4}
\usepackage{graphicx}
\usepackage{subfigure}
\usepackage[dvipdfm]{hyperref}

\begin{document}

\title{Quantum Computation and Bell-state Measurement with Double-Dot Molecules}
\author{Hui Zhang}
\author{Guo-Ping Guo}
\email{gpguo@ustc.edu.cn}
\author{Tao Tu}
\author{Guang-Can Guo}
\affiliation{Key Laboratory of Quantum Information, University of Science and Technology
of China, Chinese Academy of Sciences, Hefei 230026, People's Republic of
China}
\date{\today}

\begin{abstract}
We propose a quantum computation architecture of double-dot
molecules, where the qubit is encoded in the molecule two-electron
spin states. By arranging the two dots inside each molecule
perpendicular to the qubit scaling line, the interactions between
neighboring qubits are largely simplified and the scaling to
multi-qubit system becomes straightforward. As an Ising-model
effective interaction can be expediently switched on and off between
any two neighboring molecules by adjusting the potential offset
between the two dots, universal two-qubit gates can be implemented
without requiring time-dependent control of the tunnel coupling
between the dots. A Bell-state measurement scheme for qubit encoded
in double-dot singlet and triplet states is also proposed for
quantum molecules arranged in this way.
\end{abstract}

 \maketitle

\section{Introduction}

\begin{figure}[tb]
  \subfigure [] {\label{fig:1:a}\includegraphics[width=0.85\columnwidth]{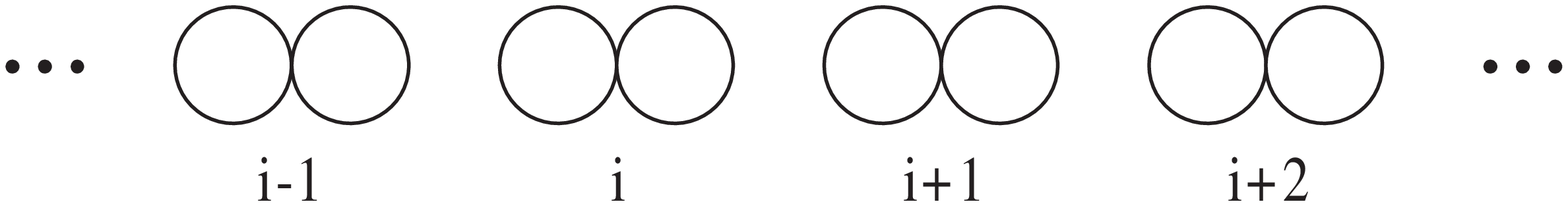}}
  \subfigure [] {\label{fig:1:b}\includegraphics[width=0.85\columnwidth]{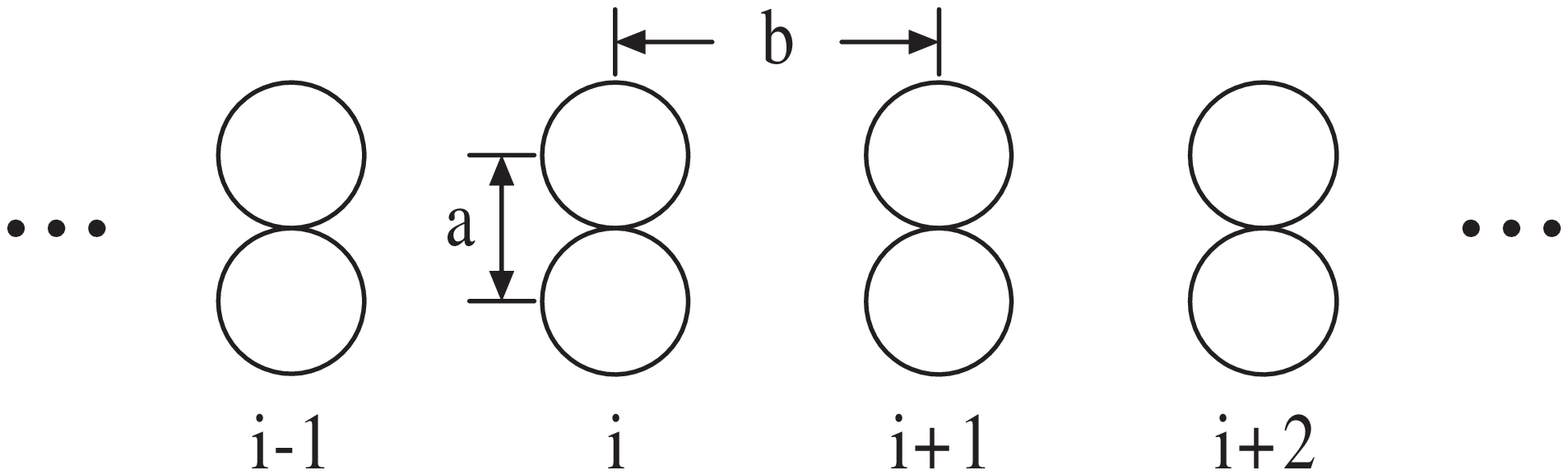}}
 \caption{Schematic diagram of double-dot qubit system. All quantum
dots have same size, and their radius is denoted by $r$. (a) The
structure that all quantum dots are arranged in line. (b) The
structure that the two dots inside each molecule perpendicular to
the qubit scaling line. The distance between two dots of each double
dot is $a$, and the distance between two double dots is
$b$.}\label{fig:1}
\end{figure}

In 1998, D. Loss and D. P. DiVincenzo proposed a quantum computation
(QC) protocol based on electron spin trapped in semiconductor
quantum dot\cite{D. Loss}. Compared with other systems such as
optics, atoms and nuclear magnetic resonance (NMR), this solid
system is argued to be more scalable and can be compatible to the
present semiconductor technology\cite{Hanson}. Recently,
two-electron spin states in double quantum dots have attracted many
interests\cite{Levy,Lidar pra,Lidar prl}. The Initialization,
manipulation and detection of these two-electron spin states have
been theoretically analyzed and experimentally
demonstrated\cite{J.R. Petta,F.H.L. Koppens,J.M. Taylor}. Then there
is the idea to encode qubit on the singlet state $S=\left(
\left\vert \uparrow \downarrow \right\rangle -\left\vert \downarrow
\uparrow \right\rangle \right) /\sqrt{2}$ and the triplet state
$T=\left( \left\vert \uparrow \downarrow \right\rangle +\left\vert
\downarrow \uparrow \right\rangle \right) /\sqrt{2}$ of double
coupled quantum dots\cite{J.M. Taylor et al.}. A fault-tolerant
architecture for QC is also proposed for these two-spin
qubits\cite{J.M. Taylor et al.}. It is argued that this encoding can
protect qubits from low-frequency noise, and suppresses the dominant
source of decoherence from hyperfine interactions\cite{J.R.
Petta,Wu,Taylor,Petta,A.C. Johnson,de Sousa}. In the papers\cite{G.
Burkard,Stepanenko}, the quantum molecules are arranged in line and
the two dots inside each molecule are also arrayed in the same line
as shown in Fig.\ref{fig:1:a}. The qubit is encoded in the singlet
and triplet states of the double dots inside each molecule and the
neighboring qubits are coupled by the direct Coulomb repulsion of
the electrons between different molecules. Two-qubit controlled-not
gate are analyzed in details when four quantum dots (two molecules)
are ranged in line.

Different to the previous one-dimensional alinement of all quantum dots\cite%
{G. Burkard,Stepanenko,J.M. Taylor et al.}, we here propose an
architecture to arrange: the two dots inside each molecule
perpendicular to the qubit scaling line as Fig.\ref{fig:1:b}. As the
qubit is encoded in the two-electron spin states of each molecule,
an Ising-model effective interaction can be switched on and off
between any two neighboring molecules without affecting other
neighboring qubits in this architecture. Universal two-qubit gates
can be implemented without requiring time-dependent control of the
tunnel coupling between the dots. A Bell-state measurement for qubit
encoded in double-dot singlet and triplet states is also analyzed
for quantum molecules arranged in this way.

In the section \ref{section:initialization}, we analyze the qubit
initialization and single qubit operations. The realization of
two-qubit gate operations is detailedly investigated in section III
. Section IV includes the single qubit readout and Bell state
measurement. In the last section, we give some discussions and
present our conclusion.
\begin{figure}[tb]
  \includegraphics[width=0.85\columnwidth]{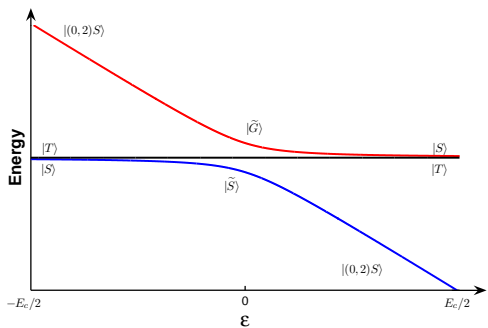}
 \caption{Energy level structure of the double dot system. The range of
the bias voltage $\varepsilon $ which we interested in is between
$-E_{c}/2$ and $E_{c}/2$.}\label{fig:2}
\end{figure}

\section{Initialization and single qubit
operations}\label{section:initialization}

Consider two double-dot molecules as shown in Fig.\ref{fig:1:b}. The
two dots inside each molecule perpendicular to the qubit scaling
line. Inside each quantum molecule, there are generally three energy
favorable states $\left\vert \left( 1,1\right) S\right\rangle $,
$\left\vert \left( 1,1\right) T\right\rangle $ and $\left\vert
\left( 0,2\right) S\right\rangle $ included for coupled double-dot
due to Coulomb blockade and Pauli blockade\cite{Hanson2,J.R.
Petta,F.H.L. Koppens,J.M. Taylor}. The notation $\left(
n_{u},n_{l}\right) $ indicates $n_{u}$ electrons on the ``upper''
dot of each qubit and $n_{l}$ electrons on the ``lower'' dot. Define
a parameter $\varepsilon $ to represent potential offset between two
dots inside each molecule, which can be changed by external
electrical field or bias voltage on the gates defining quantum
dots\cite{G.P. Guo and H. Zhang}. The range of $\varepsilon $ can be
changed between $-E_{c}/2$ and $E_{c}/2$, and for this case, there
are only two charge states of each qubit: $\left( 0,2\right) $ and
$\left( 1,1\right) $. Here $E_{c}$ indicates the charging energy of
each dot.\ Due to Pauli blockade, the double-dot state can shift to
charge state $\left( 0,2\right) $ if the initial state is
$\left\vert(1, 1) S\right\rangle $ (we use $\left\vert
S\right\rangle$ to denote $\left\vert(1, 1) S\right\rangle $ in the
following text), but remain in charge state $\left( 1,1\right) $ if
the initial state is $\left\vert T\right\rangle $. The energy level
structure of each molecule is shown in Fig.\ref{fig:2}. Due to the
tunneling between the two dots, the charge states $\left(
0,2\right) $ and $\left( 1,1\right) $ hybridize. According to reference \cite%
{J.M. Taylor}, we can define two superposition states
\begin{equation}
\begin{aligned} \left\vert \widetilde{S}\right\rangle &=&\cos \theta
\left\vert S\right\rangle +\sin \theta \left\vert ( 0,2) S\right\rangle , \\
\left\vert \widetilde{G}\right\rangle &=&-\sin \theta \left\vert
S\right\rangle +\cos \theta \left\vert ( 0,2) S\right\rangle . \end{aligned}
\end{equation}%
where
\begin{equation}
\theta =\arctan \left( \frac{2T_{c}}{\varepsilon -\sqrt{4\left\vert
T_{c}\right\vert ^{2}+\varepsilon ^{2}}}\right) .  \label{equ:theta}
\end{equation}%
Here $T_{c\text{ }}$ indicates the tunnel coupling. By adiabatically
sweeping $\varepsilon $ from $-E_{c}/2$ to $E_{c}/2$, the double-dot states $%
\left\vert \widetilde{S}\right\rangle $ and $\left\vert \widetilde{G}%
\right\rangle $ evolve according to Eq. 1. The rapid adiabatic passage means
that the transformation of $\varepsilon $ is fast relative to the nuclear
mixing time $\sim \hbar /\left( g^{\ast }\mu _{B}B_{nuc}\right) $ but slow
with respect to the tunnel coupling $T_{c}$\cite{J.R. Petta,G.P. Guo and H.
Zhang}. Here $g^{\ast }$ is the effective $g$-factor of the electron and $%
\mu _{B}$ is the Bohr magneton. When $\varepsilon =-E_{c}/2\ll -\left\vert
T_{c}\right\vert $, the adiabatic angle $\theta \rightarrow 0$, and the
eigenstate $\left\vert \widetilde{S}\right\rangle \rightarrow \left\vert
 S\right\rangle $, $\left\vert \widetilde{G}\right\rangle
\rightarrow \left\vert \left( 0,2\right) S\right\rangle $. When $\varepsilon
=E_{c}/2$, we get $\theta \rightarrow \pi /2$ and the eigenstates $%
\left\vert \widetilde{S}\right\rangle \rightarrow \left\vert
(0,2)S\right\rangle $, and $\left\vert \widetilde{G}\right\rangle
\rightarrow \left\vert S\right\rangle $. Thus by adiabatically
sweeping $\varepsilon $ from $-E_{c}/2$ to $E_{c}/2$, the double
dots initially in the state $\left\vert S\right\rangle $ can change
from the charge state $(1,1)$ to $(0,2)$. For simplicity, we use
$\left\vert S^{\prime }\right\rangle $ to represent the singlet
state $\left\vert \left( 0,2\right) S\right\rangle $ in the
following text.

We can initialize the system to the state $\left\vert S\right\rangle
$ by loading two electrons from a nearby Fermi sea into the ground
state of a single quantum dot $\left\vert S^{\prime }\right\rangle $
and then sweeping the bias voltage $\varepsilon $ from $E_{c}/2$ to
$-E_{c}/2$ in the rapid
adiabatic passage to change the charge state from $(0,2)$ to $(1,1)$\cite%
{J.R. Petta,J.M. Taylor,J.M. Taylor et al.}. In the following, we can see
that when the neighboring qubits are both in the charge state $(0,2)$, the
interaction between them will be switched on. Thus, the initialization can
only be simultaneously made on non-neighboring qubits. For a one-dimensional
qubit array, we need at least two steps to initialize all qubit to the $(1,1)
$ charge state $\left\vert S\right\rangle $. For a two-dimensional array,
four steps are needed. After initialization, the bias voltage $\varepsilon $
of all qubits are kept in the value of $-E_{c}/2$ and all qubit charge
states are in $(1,1)$.

According to the Euler angle method, if rotations by arbitrary
angles about two orthogonal axes are available, arbitrary
single-qubit rotations can be constructed. For the present
double-dot molecule, it has also been analyzed and experimentally
shown that arbitrary single-qubit rotations can be performed at
finite singlet-triplet energy splitting $J$, by combining $Z$
rotations $U_{Z}$ with rotations $U_{XZ}$ around an axis in the $XZ$
plane. For example, a rotation about the $X$-axis can be generated
by a three-step sequence $U_{X}=U_{XZ}U_{Z}U_{XZ}$\cite{J.M. Taylor
et al.,G. Burkard}.

\section{Two-qubit gate}

As the qubit is encoded in the $(1,1)$ charge singlet state
$\left\vert S\right\rangle $ and triplet state $\left\vert
T\right\rangle $, we can switch on the interaction between
neighboring qubits by simultaneously changing them to charge state
$(0,2)$ only when they are both initially in the singlet state
$\left\vert S\right\rangle $. Assume that $\varepsilon =-E_{c}/2$
and each molecule is initialized in $\left( 1,1\right) $ charge
state. The Coulomb interaction between the two nearest-neighbor
qubits can be directly described by the Hamiltonian:
\begin{equation}
H_{int}=\mathrm{diag}\left\{
H_{int_{0}},H_{int_{0}},H_{int_{0}},H_{int_{0}}\right\}
\end{equation}%
in the basis $\left\vert TT\right\rangle ,$ $\left\vert T\widetilde{S}%
\right\rangle ,$ $\left\vert \widetilde{S}T\right\rangle $ and $\left\vert
\widetilde{S}\widetilde{S}\right\rangle ,$where

\begin{equation}
H_{int_{0}}=\frac{1}{4\pi \epsilon }\left( \frac{2e^{2}}{b}+\frac{2e^{2}}{%
\sqrt{a^{2}+b^{2}}}\right) .
\end{equation}%
Here $\epsilon $ is dielectric constant of GaAs, $a$ is the distance between
the dots inside each molecule and $b$ is the distance between neighboring
molecules.

When $\varepsilon $ is adiabatically swept from $-E_{c}/2$ towards $E_{c}/2$%
, the double-dot state initially in singlet state $\left\vert
S\right\rangle $ will evolve as $\left\vert
\widetilde{S}\right\rangle $ of Eq. 1. The triplet state $\left\vert
T\right\rangle $ will remain unchanged in the charge state $(1,1)$.
When $\varepsilon =E_{c}/2$, the state $\left\vert S\right\rangle $
evolves into $(0,2)$ charge state $\left\vert S^{\prime
}\right\rangle $. Then the interaction between these two molecules
can be written as:
\begin{equation}
H_{int}^{\prime }=\mathrm{diag}\left\{
H_{int_{0}},H_{int_{0}},H_{int_{0}},H_{S^{\prime }S^{\prime }}\right\}
\end{equation}%
in the basis $\left\vert TT\right\rangle ,$ $\left\vert T\widetilde{S}%
\right\rangle ,$ $\left\vert \widetilde{S}T\right\rangle ,$ $\left\vert
\widetilde{S}\widetilde{S}\right\rangle $, where $H_{S^{\prime }S^{\prime
}}= $ $e^{2}/\left( \pi \epsilon b\right) $.

Eliminating a constant background interaction $H_{int_{0}}$, we get
an effective two-molecule interaction:
\begin{equation}
\Delta H_{int}=H_{int}^{\prime }-H_{int}=\mathrm{diag}\left\{
0,0,0,H_{cc}\right\} ,
\end{equation}%
which can be switched on by sweeping $\varepsilon $ from $-E_{c}/2$ towards $%
E_{c}/2$. Here $H_{cc}=H_{S^{\prime }S^{\prime }}-$ $H_{int_{0}}$
can be regarded as the differential cross-capacitance energy between
the two double-dot systems. It is noted that the effective
interaction is switched on whenever the state $ \left\vert
\widetilde{S}\right\rangle$ includes the component of $(0,2)$ charge
state\cite{G.P. Guo and H. Zhang}. The differential
cross-capacitance energy $H_{cc}$ can thus be written as a function
of $\theta $:

\begin{equation}
H_{cc}=\frac{\left\vert \sin \theta \right\vert ^{2}}{4\pi \epsilon }(\frac{%
2e^{2}}{b}-\frac{2e^{2}}{\sqrt{a^{2}+b^{2}}}).
\end{equation}%
When $\varepsilon =-E_{c}/2,$ $\theta \rightarrow 0$ and $H_{cc}$ is off.
When $\varepsilon =E_{c}/2$, $\theta \rightarrow \pi /2$ and and $H_{cc}$ is
maximal.

Combined with some single qubit operations, which have been shown
available for double-dot molecule in
Section.\ref{section:initialization}, we can construct any two-qubit
gate and realize universal quantum computation with the present
Ising-model effective interaction. For example the controlled-not
gate can be achieved with two single-qubit Hadamard operations
$\sigma _{H}$ and a two-qubit operation
$U(t_{0})=\mathrm{diag}\left\{ 1,1,1,-1\right\} $ as

\begin{equation}
U_{C-NOT}=\left\{ I_{1}\otimes \sigma _{H_{2}}\right\} U(t_{0})\left\{
I_{1}\otimes \sigma _{H_{2}}\right\} .
\end{equation}%
By choosing a proper interaction time $t=t_{0}$ for $H_{cc}t/\hbar =\pi
,3\pi ,5\pi ...$, we can get the two-qubit operation $U(t_{0})$ directly
from the present effective interaction:

\begin{equation}
U(t)=\exp \left( \frac{i\Delta H_{int}t}{\hbar }\right) .
\end{equation}%
After interaction time $t_{0}$, the $\varepsilon $ should be in the value of
$-E_{c}/2$ to completely switch off the effective interaction.

Comparing with the one-dimensional alinement of all quantum dots,
the present two-dimensional architecture can greatly simplify the
interaction between the neighboring quantum molecules as
Fig.\ref{fig:1:b}: there is effective interaction only when the two
neighboring molecules are both in the charge state $(0,2)$. We can
switch on the interaction between any two neighboring qubits (qubit $i$ and $%
i+1$) by simultaneously changing their charge state from $(1,1)$ to
$(0,2)$. Other neighboring qubits such as qubit $i-1$ and $i+2$ are
kept in the charge state $(1,1)$ so that they can not be infected by
the operations on qubit $i$ and $i+1$. It is noted the Coulomb
interaction between two electrons inside each qubit can also be
neglected. Only the interaction between the nearest-neighbor
molecules is included in the previous protocols. It can be shown
that the interaction between non-nearest-neighbor qubits can be
neglected safely \cite{G. Burkard,Stepanenko,G.P. Guo and H.
Zhang,X.J. Hao}.

If all the quantum dots are arranged in line as Fig.\ref{fig:1:a},
we can get an effective interaction between neighboring quantum
molecules by sweeping the two logical qubits into charge state
$(0,2)$ and $(2,0)$ respectively. In this case, the notation
$(n_{L}, n_{R})$ indicates $n_{L}$ electrons on the ``left'' dot of
each qubit and $n_{R}$ electrons on the ``right'' dot. In the basis
$\left\vert TT\right\rangle ,$ $\left\vert T\widetilde{S}
\right\rangle ,$ $\left\vert \widetilde{S}T\right\rangle ,$
$\left\vert \widetilde{S}\widetilde{S}\right\rangle $, the
interaction between two neighboring qubits $i$ and $i+1$ can be
written in the form:
\begin{equation}
H_{int}^{\prime }=\mathrm{diag}\left\{ H_{int_{0}}^{\prime
},H_{int_{0}}^{\prime }+E,H_{int_{0}}^{\prime }+E,H_{int_{0}}^{\prime
}+E^{\prime }\right\} ,
\end{equation}
where $H_{int_{0}}^{\prime }$ is the interaction between two quantum
molecules which are both in the $(1,1)$ charge states; $E$ is the
Coulomb interaction energy change when qubit $i$ (or $i+1$) is swept
from the charge state $(1,1)$ to $(0,2)$ (or $(2,0)$); $E^{\prime }$
is the energy change when qubit $i$ and $i+1$ are swept from the
charge state $(1,1)$ to $ (0,2)$ and $(2,0)$ respectively. When
there are only two logical qubits, this interaction can be used to
get two-qubit gates \cite{G. Burkard, Stepanenko} . However, for the
scalable quantum computation, we can not exclusively switch on
interaction between two neighboring qubits without influencing other
neighboring qubits. For example, when we switch on interaction $
H_{int}^{\prime }$ between qubit $i$ and $i+1$, there is also an
effective interaction between qubit $i-1$ and $i$ (or qubit $i+1$
and $i+2)$:
\begin{equation}
H_{int}^{\prime \prime }=\mathrm{diag}\left\{ H_{int_{0}}^{\prime
},H_{int_{0}}^{\prime },H_{int_{0}}^{\prime }-E,H_{int_{0}}^{\prime
}-E\right\} .
\end{equation}%
As we need to change the molecule charge state in the measurement
process, the kind of unavoidable effect from other neighboring
qubits will be switched on and influence the single qubit readout
for this architecture that all quantum dots are arranged in line.
\begin{figure}[ht]
  \includegraphics[width=0.95\columnwidth]{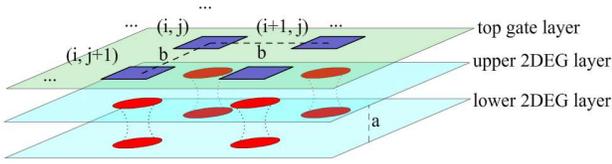}
 \caption{Schematic diagram of two-dimensional array when bilayer
2-dimensional-electron-gas (2DEG) are used to form quantum molecule.
The squares in the top gate layer and circles in the two 2DEG layers
stand for controlling gates and quantum dots
respectively.}\label{fig:3}
\end{figure}

Although the quantum dots are arranged in two-dimensional
architecture, we have only considered a one-dimensional logical
qubits or quantum molecules chain in the above discussion. Actually,
we can scale the logical qubits to a two-dimensional array when
bilayer 2-dimensional-electron-gas (2DEG) are used to form quantum
molecule\cite{macdonald,dassarma,X.J. Hao} as shown in
Fig.\ref{fig:3}. Each molecule comprises one quantum dot in the
upper 2DEG and another one in the lower 2DEG. The two quantum dots
of upper and lower layers are tunneling coupled to form one quantum
molecules. These molecules can also be the self-assembled quantum
dot pillars, which also include two quantum dots respectively in the
upper and lower part of the pillar\cite{yamamoto,X.J. Hao}.

\section{Single qubit readout and Bell state measurement}
\begin{figure}[tb]
  \includegraphics[width=0.65\columnwidth]{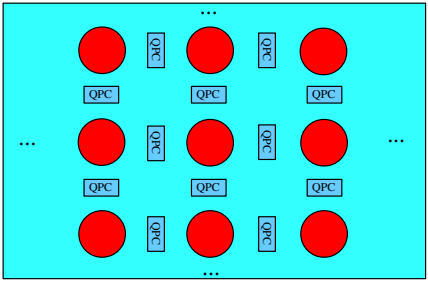}
 \caption{The structure diagram of single-qubit and two-qubit
measurement in the lower 2DEG layer. The quantum point contact (QPC)
is used to detect the distribution of two electrons in the nearby
double-dot molecule.}\label{fig:4}
\end{figure}

As the $(1,1)$ charge state $\left\vert S\right\rangle $ and the
$(0,2)$ charge state $\left\vert S^{\prime }\right\rangle $ can
transform mutually by adjusting the bias voltage $\varepsilon $, the
single qubit readout for the present double-dot molecule can be made
by a quantum point contact (QPC) placed near one of the quantum dot
as shown in Fig.\ref{fig:4} \cite{J.R. Petta, J.M. Taylor}. From the
current through QPC, we can know the two electrons distribution in
the double-dot molecule. If the molecule is in the singlet
state $\left\vert S\right\rangle $, it will be changed to the charge state $%
(0,2)$ and the current of QPC will be lower, when we sweep $\varepsilon $
from $-E_{c}/2$ to $E_{c}/2$. If the molecule is in the triplet state $%
\left\vert T\right\rangle $, it will stay in the charge state $(1,1)$ and
the current of QPC will be unchanged when sweeping $\varepsilon $. As the
interaction between neighboring qubits will be switched on when they are
both in the charge state $(0,2)$, we will avoid to perform single qubit
measurement on two neighboring molecules simultaneously.

With the present architecture of quantum molecules, we can also
perform Bell state measurement for these qubits encoded in
double-dot singlet and triplet states, by directly place a QPC in
the middle of two molecules as shown in Fig.\ref{fig:4}. Any
two-qubit state can be denoted by:
\begin{equation}
\Phi _{12}=p_{1}\Phi ^{+}+p_{2}\Phi ^{-}+p_{3}\Psi ^{+}+p_{4}\Psi ^{-},
\end{equation}%
where $p_{1},p_{2},p_{3},p_{4}\in C$, $\left\vert p_{1}\right\vert
^{2}+\left\vert p_{2}\right\vert ^{2}+\left\vert p_{3}\right\vert
^{2}+\left\vert p_{4}\right\vert ^{2}=1$. $\Phi ^{\pm }=\left(
\left\vert TT\right\rangle \pm \left\vert SS\right\rangle \right)
/\sqrt{2}$ and $\Psi ^{\pm }=\left( \left\vert TS\right\rangle \pm
\left\vert ST\right\rangle \right) /\sqrt{2}$ are the four Bell
states. We can detect the charge state of the quantum dots through
the QPC current $I$. Simultaneously sweep the bias voltage $
\varepsilon $ of the two qubits, which need to be measured, from $-E_{c}/2$ to $%
E_{c}/2$. Due to the distribution of the four electrons in the two
molecules, the QPC current $I$ can thus have three different values:
the current $I$ is kept in the value $I_{\max }$, which means the
two qubits are both in the charge state $(1,1)$; the current $I$ is
changed to
the value $I_{\min }$, which means that both qubits are in the charge state $%
(0,2)$; the current $I$ gets a value $I_{mid}$ smaller than $I_{\max }$ but
larger than $I_{\min }$, which corresponds to the case that one of the two
qubits is in the charge state $(0,2)$.

As the effective interaction will be switched on when the two qubits
are both in the charge state $(0,2)$, the two-qubit state $\Phi
_{12}$ will evolve as
in the following form when sweeping $\varepsilon $ from $-E_{c}/2$ to $%
E_{c}/2:$
\begin{eqnarray}
\Phi _{12}^{\prime } &=&\frac{p_{1}}{\sqrt{2}}\left( \left\vert
TT\right\rangle +e^{i\varphi
}\left\vert \widetilde{S}\widetilde{S}\right\rangle \right) %
+\frac{p_{2}}{\sqrt{2}}\left( \left\vert TT\right\rangle -e^{i\varphi }\left\vert \widetilde{S%
}\widetilde{S}\right\rangle \right)  \nonumber \\
&&+\frac{p_{3}}{\sqrt{2}}\left( \left\vert T\widetilde{S}\right\rangle +\left\vert \widetilde{%
S}T\right\rangle \right)+\frac{p_{4}}{\sqrt{2}}\left( \left\vert T\widetilde{S}%
\right\rangle -\left\vert \widetilde{S}T\right\rangle \right)
\end{eqnarray}%
where $\varphi =\int_{0}^{t_{m}}H_{cc}dt$, $t_{m}$ represents the time that $%
\varepsilon $ leaves the value $-E_{c}/2$. The effective interaction only
adds a phase to the component $\left\vert \widetilde{S}\widetilde{S}%
\right\rangle $. As it can only be switched on only when the two
qubits are both in the state $\left\vert SS\right\rangle $, it has
no influence to $\Psi ^{\pm }$. If the QPC current $I$ gets the
value $I_{\max }$, the two qubits are thus both in the charge state
$(1,1)$. This means that the two qubits are in the state $\left\vert
TT\right\rangle $. The current $I_{\min }$ means that the two qubits
are both in the charge state $(0,2)$. Then we can determine that
the two-qubit state is in the state $\left\vert \widetilde{S}\widetilde{S}%
\right\rangle $, which is evolved from the initial state $\left\vert
SS\right\rangle $.

If the QPC current $I=I_{mid}$, we can know that the one of the two
qubits is in the charge state $(0,2)$. This is the case that the two
qubits are in the state $\left( \left\vert
T\widetilde{S}\right\rangle +\left\vert \widetilde{S}T\right\rangle
\right) /\sqrt{2}$ or $\left( \left\vert T\widetilde{S}\right\rangle
-\left\vert \widetilde{S} T\right\rangle \right) /\sqrt{2}$, which
respectively evolve from the initial two-qubit state $\Psi ^{+}$ or
$\Psi ^{-}$. With this step of QPC measurement, we can get the
parity information of these two qubits. In the case of $ I=I_{mid}$,
we sweep the $\varepsilon $ of the both two qubits back to $
-E_{c}/2$ from $E_{c}/2$ and then perform Hadamard operation on the
two qubits in turn. It is ensured that the two qubits can not be
simultaneously both in the charge state $(0,2)$ in the operations.
The two Hadamard operations will rotate $\Psi ^{\pm }$ respectively
into the state $\Phi ^{-}$ and $-\Psi ^{-}$\cite{Guo}. Sweep the
$\varepsilon $ of the both two qubits from $-E_{c}/2$ to $E_{c}/2$
and make QPC measurement again. As the above measurement, the QPC
current $I^{\prime }$ can also have three different values: $I_{\max
}$, $I_{mid}$, $I_{\min }$. From the values of $I$ and $ I^{\prime
}$, we can determine two Bell state $\Psi ^{+}$ and $\Psi ^{-}$ as
shown in Table.\ref{table}. We can sweep the molecule charge state
back to $(1,1)$ after this measurement. It is noted that this Bell
state measurement is not a completed one, and we can only
distinguish two of the four Bell states. In addition, this
measurement can also be regarded as a processing of generating Bell
state $\Psi ^{-}$, as the two quantum molecules which are measured
in this Bell state can be used in future applications.

\begin{table}[ht]
\begin{center}
\begin{tabular}{|c|c|c|}
\hline & $I$ & $I^{\prime }$ \\ \hline $\Phi ^{+}$ & $I_{\max}$ or
$I_{\min }$ & $-$ \\ \hline $\Phi ^{-}$ & $I_{\max }$ or $I_{\min }$
& $-$ \\ \hline $\Psi ^{+}$ & $I_{mid}$ & $I_{\max }$
or $I_{\min }$ \\ \hline $\Psi ^{-}$ & $I_{mid}$ & $I_{mid}$\\ \hline%
\end{tabular}
\caption{The states of the current through the QPC corresponding to
each Bell state.}\label{table}
\end{center}
\end{table}

\section{Discussion and conclusion}

By encoding in singlet and triplet states, qubits are protected from
low-frequency noise and the effect of homogeneous hyperfine interactions for
double dots. Recent experiments have demonstrated that the coherence time of
the singlet and triplet states can be about $10ns$, which can even be
increased to $1\mu s$ with spin-echo techniques\cite{J.R. Petta,J.M. Taylor}%
. As the rapid adiabatic passage of $\varepsilon $ is required to be
fast relative to the nuclear mixing time $\sim \hbar /\left( g^{\ast
}\mu _{B}B_{nuc}\right) $ but slow with respect to the tunnel
coupling $T_{c}\sim 0.01meV$, the $\varepsilon $ sweeping speed is
about $5meV/ns$ in these experiments. If the quantum dot of molecule
has a diameter of $100nm$, the charge energy $E_{c}\sim 5meV$ and
sweeping $\varepsilon $ from $-E_{c}/2$ to $E_{c}/2$ needs a time of
about $1ns$. For quantum molecules based on bilayer 2DEG as
Fig.\ref{fig:3}, the distance between the double dots of each
molecules $a=20nm$ and the distance between neighboring molecules
$b=10a=200nm$ (in order to safely neglect the effect from the
interaction between non-nearest-neighbor qubits), and we need a time
of about $1ns$ to achieve a two-qubit controlled phase operation
$U=\mathrm{diag}\left\{ 1,1,1,-1\right\} $\cite{G.P. Guo and H.
Zhang}. Actually, in the the previous protocols of arranging four
quantum dots of two molecules in line, qubits are similarly coupled
by Coulomb interaction\cite{G. Burkard, Stepanenko}. The two-qubit
operations may also need a time of about $1ns$. Therefore, we still
need to increase the coherence time or increase the interaction
strength, even qubit is encoded in singlet and triplet states for
these quantum computation schemes exploring Coulomb interaction to
realizing two-qubit gates. Since the QPC measurement needs a time of
about $1\mu s$, QPC measurement can be implemented only once within
the coherence time. Thus only partial Bell state measurement for
qubit encoded in singlet and triplet state may be realizable with
the present experiment conditions. Generally, we can also
distinguish the four Bell states by firstly transfer them into four
product states respectively and then simultaneously performing QPC
measurement on each qubit within the coherence time.

In conclusion, we have proposed a quantum computation architecture
based on double-dot quantum molecules. As the qubit is encoded in
the $(1,1)$
charge singlet state $\left\vert S\right\rangle $ and triplet state $%
\left\vert T\right\rangle $, we can simplify the Coulomb interaction
to a switchable Ising interaction in the present architecture.
Compared with the previous schemes, the effective Ising interaction
can be switched on and off between any two neighboring qubits
without affecting other neighboring qubits. A Bell-state measurement
scheme is also presented for qubit encoded in the singlet and
triplet state. Universal quantum gates can be performed by only
tuning the potential offset between the two dots of each molecule,
where the time-dependent control of the tunnel coupling between the
dots is eliminated.

This work was funded by National Fundamental Research Program, the
innovation funds from Chinese Academy of Sciences and National Natural
Science Foundation of China (Grant No.60121503 and No.10604052).

\end{document}